\documentclass[conference]{IEEEtran}

\usepackage{cite}
\usepackage{graphicx}
\usepackage{amsmath}  
\usepackage{algorithm,algpseudocode}
\usepackage{algorithmicx}
\usepackage{algpseudocode}
\usepackage{algorithm}
\usepackage{color}
\usepackage{verbatim}
\usepackage{amsfonts}
\usepackage{amsmath,amssymb}
\usepackage{url}
\usepackage{bbm}
\usepackage{bm}
\usepackage{subfig}
\usepackage{soul}
\usepackage{stmaryrd}
\usepackage{setspace}

\DeclareMathOperator*{\maximize}{maximize}
\DeclareMathOperator{\subjectto}{subject~to}
\newcommand{\norm}[1]{\left\lVert#1\right\rVert}

\makeatletter
\def\endthebibliography{%
	\def\@noitemerr{\@latex@warning{Empty `thebibliography' environment}}%
	\endlist
}
\makeatother


\IEEEoverridecommandlockouts
\begin{document}
	\title{Deep Actor-Critic Learning for Distributed Power Control in Wireless Mobile Networks}
	\author{Yasar Sinan Nasir and Dongning Guo\\
		Department of Electrical and Computer Engineering\\
		Northwestern University, Evanston, IL 60208.
		\thanks{This material is based upon work supported by the National Science Foundation under Grants No.~CCF-1910168 and No.~CNS-2003098 as well as a gift from Intel Incorporation.}
	}
	\maketitle
	
	\begin{abstract}
		Deep reinforcement learning offers a model-free alternative to supervised deep learning and classical optimization for solving the transmit power control problem in wireless networks. The multi-agent deep reinforcement learning approach considers each transmitter as an individual learning agent that determines its transmit power level by observing the local wireless environment. Following a certain policy, these agents learn to collaboratively maximize a global objective, e.g., a sum-rate utility function. This multi-agent scheme is easily scalable and practically applicable to large-scale cellular networks. In this work, we present a distributively executed continuous power control algorithm with the help of deep actor-critic learning, and more specifically, by adapting deep deterministic policy gradient. Furthermore, we integrate the proposed power control algorithm to a time-slotted system where devices are mobile and channel conditions change rapidly. We demonstrate the functionality of the proposed algorithm using simulation results.
	\end{abstract}
	\section{Introduction}
	With ever-increasing number of cellular devices, interference management has become a key challenge in developing newly emerging technologies for wireless cellular networks. An access point (AP) may increase its transmit power to improve data rate to its devices, but this will cause more interference to nearby devices. Power control is a well-known interference mitigation tool used in wireless networks. It often maximizes a non-convex sum-rate objective. It becomes NP-hard when multiple devices share a frequency band \cite{Luo2008dynamicspectrum}.
	
	Various state-of-the-art optimization methods have been applied to power control such as fractional programming (FP) \cite{shen2018fractional} and weighted minimum mean square error (WMMSE) algorithm \cite{sun2017learning} which are model-driven and require a mathematically tractable and accurate model \cite{qin2019DLphysical}. FP and WMMSE are iterative and executed in a centralized fashion, neglecting the delay caused by the feedback mechanism between a central controller and APs. Both require full channel state information (CSI), and APs need to wait until centralized controller sends the outcome back over a backhaul once iterations converge. 
	
	Data-driven methods are promising in a realistic wireless context where varying channel conditions impose serious challenges such as imperfect or delayed CSI. Reference \cite{sun2017learning} uses a deep neural network to mimic an optimization algorithm that is trained by a dataset composed of many optimization runs. The main motivation in \cite{sun2017learning} is to reduce the computational complexity while maintaining a comparable sum-rate performance with WMMSE. However, the training dataset relies on model-based optimization algorithms. In this paper, we consider a purely data-driven approach called model-free deep reinforcement learning. 
	
	Similar to this work, we have earlier proposed a centralized training and distributed execution framework based on deep Q-learning algorithm for dynamic (real-time) power control \cite{nasir2019deep}. Since Q-learning applies only to discrete action spaces, transmit power had to be quantized in \cite{nasir2019deep}. As a result, the quantizer design and the number of levels, i.e., number of possible actions, have an impact on the performance. For example, an extension of our prior work shows that quantizing the action space with a logarithmic step size gives better outcomes than that of a linear step size \cite{meng2019ddpgpower}. 
	
	In this work, we replace deep Q-learning with an actor-critic method called deep deterministic policy gradient (DDPG) \cite{lillicrap2015ddpg} algorithm that applies to continuous action spaces. A distributively executed DDPG scheme has been applied to power control for fixed channel and perfect CSI \cite{meng2019ddpgpower}. To the best of our knowledge, we are the first to study actor-critic based dynamic power control that involves mobility of cellular devices. Our prior work assumed immobile devices where the large-scale fading component was the steady state of the channel. We adapt our previous approach to make it applicable to our new system model that involves mobility where channel conditions vary due to both small and large scale fading. In order to ensure the practicality, we assume delayed and incomplete CSI, and using simulations, we compare the sum-rate outcome with WMMSE and FP that have full perfect CSI.
	\section{System Model and Problem Formulation}\label{sec:systemmodel}
	In this paper, we consider a special case where $N$ \emph{mobile} devices are uniformly randomly placed in $K$ homogeneous hexagonal cells. This deployment scenario is similar to the interfering multiaccess channel scenario which is also examined in \cite{sun2017learning,nasir2019deep}. Let $\mathcal{N}=\left\{1,\dots,N\right\}$ and $\mathcal{K}=\left\{1,\dots,K\right\}$ denote the sets of link and cell indexes, respectively. Here we are not concerned with the device association problem. As device $n \in \mathcal{N}$ is inside cell $k \in \mathcal{K}$, its associated AP $n$ is located at the center of cell $k$. We denote the cell association of device $n$ as $b_{n}\in \mathcal{K}$ and its AP $n$ is positioned at the center of $b_{n}$.
	
	All transmitters and receivers use a single antenna and we consider a single frequency band with flat fading. The network is assumed to be a fully synchronized time slotted system with slot duration $T$. We employ a block fading model to denote the downlink channel gain from a transmitter located at the center of cell $k$ to the receiver antenna of device $n$ in time slot $t$ as
	\begin{align}\label{eq:channel}
	\bar{g}^{(t)}_{k\to n} &= \left| h^{(t)}_{k\to n}\right|^2 \alpha^{(t)}_{k\to n}, \quad t=1,2,\dots \,.
	\end{align}
	In \eqref{eq:channel}, $\alpha^{(t)}_{k\to n} \geq 0$ represents the large-scale fading component including path loss and log-normal shadowing which varies as mobile device $j$ changes its position. Let $\mathbf{x}_k$ denote the 2D position, i.e., $(x,y)$-coordinates, of cell $k$'s center. Similarly, we represent the location of mobile device $n$ at slot $t$ as $\mathbf{x}^{(t)}_{n}$. Then, the large-scale fading can be expressed in dB as
	\begin{align}\label{eq:largefading}
	\alpha_{\textrm{dB},k\to n}^{(t)} &= \textrm{PL}\left(\mathbf{x}_k,\mathbf{x}^{(t)}_{n}\right) + \mathcal{X}^{(t)}_{k\to n},
	\end{align}
	where $\textrm{PL}$ is the distance-dependent path loss in dB and $\mathcal{X}^{(t)}_{k\to n}$ is the log-shadowing from $\mathbf{x}_k$ to $\mathbf{x}^{(t)}_{n}$. For each device $n$, we compute the shadowing from all $k$ possible AP positions in the network. The shadowing parameter is updated by $\mathcal{X}^{(t)}_{k\to n} = \rho^{(t)}_{\textrm{s},n}\mathcal{X}^{(t)}_{k\to n} + \sigma_{\textrm{s}}e^{(t)}_{\textrm{s},k\to n}$,
	where $\sigma_\textrm{s}$ is the log-normal shadowing standard deviation and the correlation $\rho^{(t)}_{\textrm{s},n}$ is computed by $\rho^{(t)}_{\textrm{s},n}=e^{\frac{\Delta\mathbf{x}^{(t)}_{n}}{d_{\textrm{cor}}}}$
	with $\Delta\mathbf{x}^{(t)}_{n} = \norm{\mathbf{x}^{(t)}_{n}-\mathbf{x}^{(t-1)}_{n}}_2$ being the displacement of device $n$ during the last slot and with $d_{\textrm{cor}}$ being the correlation length of the environment. Note that $\mathcal{X}^{(0)}_{k\to n} \sim \mathcal{N}\left(0,\sigma_s^2\right)$ and the shadowing innovation process $e^{(1)}_{\textrm{s},k\to n},e^{(2)}_{\textrm{s},k\to n},\dots$ consists of independent and identically distributed (i.i.d.) Gaussian variables with distribution $\mathcal{N}\left(0,1-\left(\rho^{(t)}_{\textrm{s},n}\right)^2\right)$. Following \cite{haas1997mobility}, we model the change in the movement behavior of each device as incremental steps on their speed and directions.
	
	Using the Jakes fading model \cite{nasir2019deep}, we introduce the small-scale Rayleigh fading component of \eqref{eq:channel} as a first-order complex Gauss-Markov process: $h_{k\to n}^{(t)} = \rho^{(t)}_{n} h_{k\to n}^{(t-1)} + e_{k\to n}^{(t)}$,
	where $h_{k\to n}^{(0)}\sim C\mathcal{N}(0,1)$ is circularly symmetric complex Gaussian (CSCG) with unit variance and the independent channel innovation process $e_{k\to n}^{(1)},e_{k\to n}^{(2)},\dots$ consists of i.i.d. CSCG random variables with distribution $C\mathcal{N}\left(0,1-\rho^2\right)$. The correlation $\rho^{(t)}_{n}$ depends on the  $\rho=J_0(2\pi f_{d,n}^{(t)} T)$, where $J_0(.)$ is the zeroth-order Bessel function of the first kind and $f_{d,n}^{(t)}=v_n^{(t)}f_c/c$ is device $n$'s maximum Doppler frequency at slot $t$ with $v_n^{(t)}=\Delta\mathbf{x}^{(t)}_{n}/T$ being device $n$'s speed, $c=3\times 10^8$ m/s, and $f_c$ being carrier frequency.
	
	Let $b^{(t)}_{n}$ and $p^{(t)}_{n}$ denote device $n$'s associated cell and transmit power of its associated AP in time slot $t$, respectively. Hence the association and allocation in time slot $t$ can be denoted as $\bm{b}^{(t)}=\left[b^{(t)}_1,\dots,b^{(t)}_N\right]^\intercal$ and $\bm{p}^{(t)}=\left[p^{(t)}_1,\dots,p^{(t)}_N\right]^\intercal$, respectively. The signal-to-interference-plus-noise ratio at receiver $n$ in time slot $t$ can be defined as a function of the association $\bm{b}^{(t)}$ and allocation $\bm{p}^{(t)}$:
	\begin{align}\label{eq:SINR}
	\gamma_{n}^{(t)}\left(\bm{b}^{(t)},\bm{p}^{(t)}\right) &= \frac{\bar{g}^{(t)}_{b_{n}^{(t)}\to n}p_n^{(t)}}{\sum_{m \neq n}\bar{g}^{(t)}_{b_{m}^{(t)}\to n}p_m^{(t)}+\sigma^2},
	\end{align}
	where $\sigma^2$ is the additive white Gaussian noise power spectral density which is assumed to be the same at all receivers without loss of generality. Then, the downlink spectral efficiency of device $n$ at time $t$ is
	\begin{align}\label{eq:DynRate}
	\begin{split}
	C^{(t)}_n &= \log\left(1+\gamma_{n}^{(t)}\left(\bm{b}^{(t)},\bm{p}^{(t)}\right)\right).
	\end{split}
	\end{align}
	For a given association $\bm{b}^{(t)}$, the power control problem at time slot $t$ can be defined as a sum-rate maximization problem:
	\begin{align}\label{eq:DynOptProblem}
	\begin{split}
	\maximize_{\bm{p}^{(t)}} & \quad \sum_{n=1}^{N} C^{(t)}_n \\
	\subjectto & \quad 0 \leq p_n \leq P_{\textrm{max}} ,\quad n = 1, \dots, N \,,
	\end{split}
	\end{align}
	where $P_{\textrm{max}}$ is the maximum power spectral density that an AP can emit. The real-time allocator solves the problem in \eqref{eq:DynOptProblem} at the beginning of slot $t$ and its solution becomes $\bm{p}^{(t)}$. For ease of notation, throughout the paper, we use $g^{(t)}_{m\to n}=\bar{g}^{(t)}_{b^{(t)}_{m}\to n}$.
	\section{Proposed Power Control Algorithm} \label{sec:algorithm}
	\subsection{Reinforcement Learning Overview}
	A learning agent intersects with its environment, i.e., where it lives, in a sequence of discrete time steps. At each step $t$, agent first observes the state of environment, i.e., key relevant environment features, $s^{(t)} \in \mathcal{S}$ with $\mathcal{S}$ being the set of possible states. Then, it picks an action $a^{(t)} \in \mathcal{A}$, where $\mathcal{A}$ is a set of actions, following a policy that is either deterministic or stochastic and is denoted by $\mu$ with $a^{(t)} = \mu(s^{(t)})$ or $\pi$ with $a^{(t)} \sim \pi(\cdot | s^{(t)})$, respectively. As a result of this interaction, environment moves to a new state $s^{(t+1)}$ following a transition probability matrix that maps state-action pairs onto a distribution of states at the next step. Agent perceives how good or bad taking action $a^{t}$ at state $s^{(t)}$ is by a reward signal $r^{(t+1)}$. We describe the above interaction as an experience at $t+1$ denoted as $e^{(t+1)}=\left(s^{(t)},a^{(t)},r^{(t+1)},s^{(t+1)}\right)$. 
	
	Model-free reinforcement learning learns directly from these interactions without any information on the transition dynamics and aims to learn a policy that maximizes agent's long-term accumulative discounted reward at time $t$, 
	\begin{align}\label{eq:discountedreward}
	\begin{split}
	R^{(t)} &= \sum_{\tau=0}^{\infty}\gamma^{\tau} r^{(t+\tau+1)}
	\end{split},
	\end{align}
	where $\gamma \in (0,1]$ is the discount factor.
	
	Two main approaches to train agents with model-free reinforcement learning are value function and policy search based methods \cite{arulkumaran2017drlsurvey}. The well-known Q-learning algorithm is value based and learns an action-value function $Q(s,a)$. The classical Q-learning uses a lookup table to represent Q-function which does not scale well for large state spaces, i.e., a high number of environment features or some continuous environment features. Deep Q-learning overcomes this challenge by employing a deep neural network to represent Q-function in place of a lookup table. However, the action space still remains discrete which requires quantization of transmit power levels in a power control problem. Policy search methods can directly handle continuous action spaces. In addition, compared to Q-learning that indirectly optimize agent's performance by learning a value function, policy search methods directly optimize a policy which is often more stable and reliable \cite{achiam2018spinup}.  By contrast, the policy search algorithms are typically on-policy which means each policy iteration only uses data that is collected by the most-recent policy. Q-learning can reuse data collected at any point during training, and consequently, more sample efficient. Another specific advantage of off-policy learning for a wireless network application is that the agents do not need to wait for the most-recent policy update and can simultaneously collect samples while the new policy is being trained. Since both value and policy based approaches have their strengths and drawbacks, there is also a hybrid approach called actor-critic learning \cite{arulkumaran2017drlsurvey}.	
	
	Reference \cite{lillicrap2015ddpg} proposed the DDPG algorithm which is based on the actor-critic architecture and allows continuous action spaces. DDPG algorithm iteratively trains an action-value function using a critic network and uses this function estimate to train a deterministic policy parameterized by an actor network.
		
	For a policy $\pi$, the $Q$-function at state-action pair $(s,a)\in \mathcal{S}\times \mathcal{A}$ becomes
	\begin{align}\label{eq:qfunction}
	Q^{\pi}(s,a) &= \mathbb{E}_{\pi}\left[R^{(t)} \middle| s^{(t)}=s, a^{(t)}=a \right].
	\end{align}
	
	For a certain state $s$, a deterministic policy $\mu: \mathcal{S} \to \mathcal{A}$ returns action $a=\mu(s)$. In a stationary Markov decision process setting, the optimal $Q$-function associated with the target policy $\mu$ satisfies the Bellman property and we can make use of this recursive relationship as	
	\begin{align}\label{eq:bellmanqfunction}
	Q^\mu(s,a)&=\mathbb{E}\left[r^{(t+1)} + \gamma Q^\mu(s',\mu(s')) \middle|s^{(t)}=s,a^{(t)}=a \right],
	\end{align}	
	where the expectation is over $s'$ which follows the distribution of the state of the environment. As the target policy is deterministic, the expectation in \eqref{eq:bellmanqfunction} depends only on the environment transition dynamics. Hence, an off-policy learning method similar to deep Q-learning can be used to learn a Q-function parameterized by a deep neural network called critic network. The critic network is denoted as $Q_{\bm{\phi}}(s,a)$ with $\bm{\phi}$ being its parameters. Similarly, we parameterize the policy using another DNN named actor network $\mu_{\bm{\theta}}(s)$ with policy parameters being $\bm{\theta}$.
	
	Let the past interactions be stored in an experience-replay memory $\mathcal{D}$ until time $t$ in the form of $e=(s,a,r',s')$. This memory needs to be large enough to avoid over-fitting and small enough for faster training. DDPG also applies another trick called quasi-static target network approach and define two separate networks to be used in training which are train and target critic networks with their parameters denoted as $\bm{\phi}$ and $\bm{\phi}_\textrm{target}$, respectively. To train $\bm{\phi}$, at each time slot, DDPG minimizes the following mean-squared Bellman error:
	\begin{align}\label{eq:MSBEloss}
	L\left(\bm{\phi},\mathcal{D}\right) &= \mathbb{E}_{(s,a,r',s') \sim \mathcal{D}} \left[\left( y(r',s') -Q_{\bm{\phi}}\left(s,a\right)\right)^2\right]
	\end{align}
	where the target $y(r',s') = r' + \gamma Q_{\bm{\phi}_{\textrm{target}}}\left(s',\mu_{\bm{\theta}}(s')\right)$. Hence, $\bm{\phi}$ is updated by sampling a random mini-batch $\mathcal{B}$ from $\mathcal{D}$ and running gradient descent using 
	\begin{align}\label{eq:criticgradient}
	\nabla_{\bm{\phi}} \frac{1}{|\mathcal{B}|} \sum_{(s,a,r',s') \in \mathcal{B}} \left(y(r',s') - Q_{\bm{\phi}}\left(s,a\right) \right)^2.
	\end{align}
	Note that after each training iteration $\bm{\phi}_\textrm{target}$ is updated by $\bm{\phi}$.
	
	In addition, the policy parameters are updated to learn a policy $\mu_{\bm{\theta}}(s)$ which gives the action that maximizes $Q_{\bm{\phi}}(s,a)$. Since the action space is continuous, $Q_{\bm{\phi}}(s,a)$ is differentiable with respect to action and $\bm{\theta}$ is updated by gradient ascent using
	\begin{align}\label{eq:actorgradient}
	\nabla_{\bm{\theta}} \frac{1}{|\mathcal{B}|} \sum_{(s,\dots) \in \mathcal{B}} Q_{\bm{\phi}}\left(s,\mu_{\bm{\theta}}(s)\right).
	\end{align}
	To ensure exploration during training, a noise term is added to the deterministic policy output \cite{lillicrap2015ddpg}. In our multi-agent framework to be discussed next section, we employ $\epsilon$-greedy algorithm of Q-learning instead for easier tuning.
	\subsection{Proposed Multi-Agent Learning Scheme for Power Control}
	\begin{figure}
	[t]
	\centering
	\includegraphics[clip, trim=0.3cm 3.0cm 17.8cm 3.50cm,width=1.0\columnwidth]{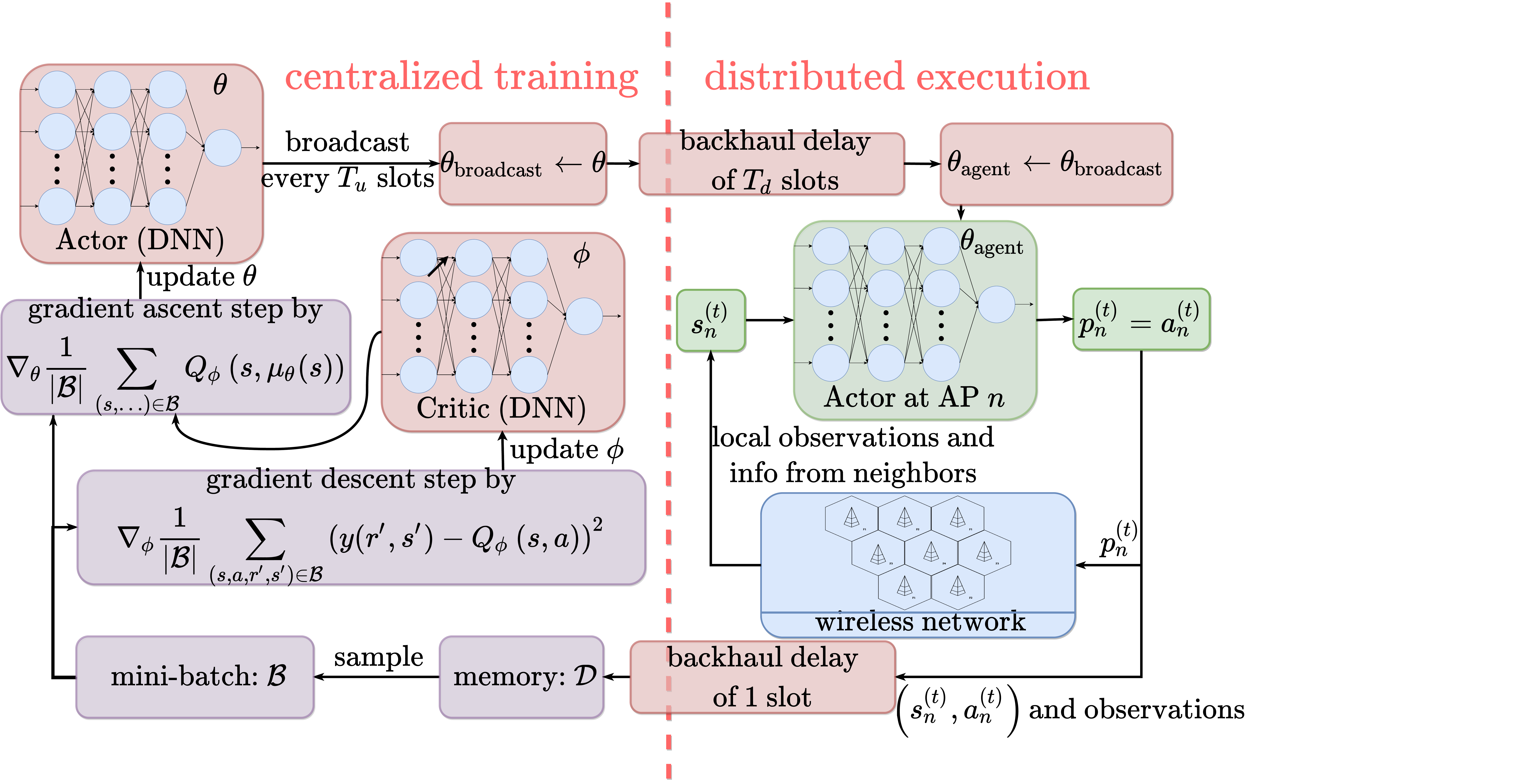}
	\caption{Diagram of the proposed power control algorithm.}
	\label{fig:ddpgdiagram}
	\end{figure}
	For the proposed power control scheme in Fig. \ref{fig:ddpgdiagram}, we let each transmitter be a learning agent. Hence, the next state of each agent is determined by the joint-actions of all agents and the environment is no longer stationary. In order to avoid instability, we gather the experiences of all agents in a single replay memory and train a global actor network $\bm{\theta}_{\textrm{agent}}$ to be shared by all agents. At slot $t$, each agent $n\in \mathcal{N}$ observes its local state $s_n^{(t)}$ and sets its own action $a_n^{(t)}$ by using $\bm{\theta}_{\textrm{agent}}$.
	
\begin{figure}
		[t]
		\centering
		\includegraphics[clip, trim=0.7cm 0.4cm 2.6cm 0.00cm,width=0.88\columnwidth]{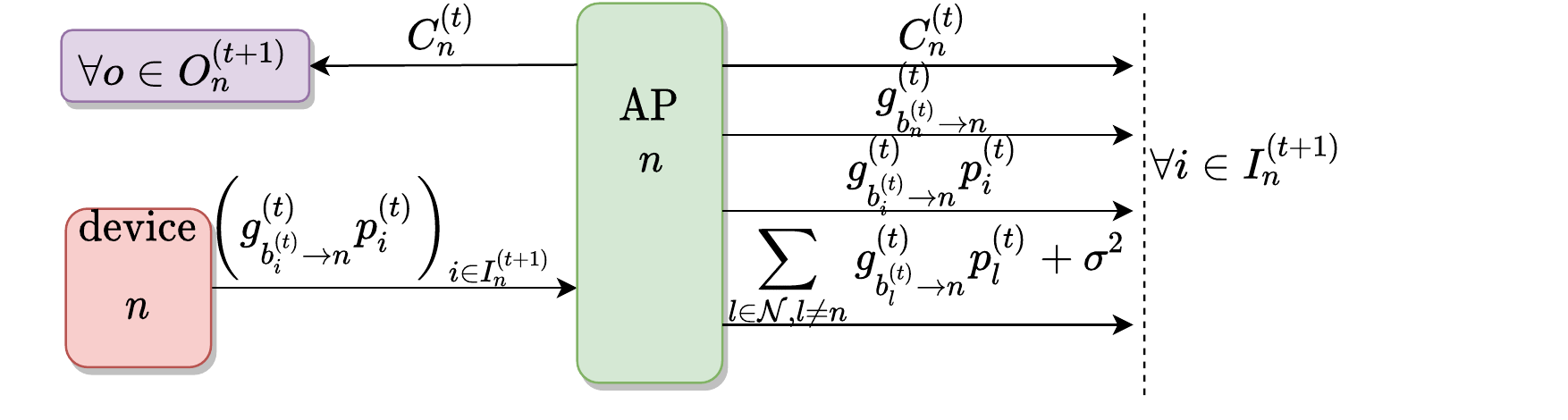}
		\caption{The information exchange in time slot $t$.}
		\label{fig:infoexchangeconf}
	\end{figure}

	For each link $n$, we first describe the neighboring sets that allow the distributively execution. Link $n$'s set of interfering neighbors at time slot $t$ consists of nearby AP indexes whose received signal-to-noise ratio (SNR) at device $n$ was above a certain threshold during the past time slot and is denoted as
	\begin{align}\label{eq:InNeigh}
	I^{(t)}_{n} = \left\{i\in \mathcal{N}, i \neq n \middle| g^{(t-1)}_{i\to n}p^{(t-1)}_i  > \eta \sigma^2 \right\}.
	\end{align}
	Conversely, we define link $n$'s set of interfered neighbors at time slot $t$ using the received SNR from AP $n$, i.e.,
	\begin{align}\label{eq:OutNeigh}
	O^{(t)}_{n} = \left\{o \in \mathcal{N}, o \neq n \middle| g^{(t-1)}_{n\to o}p^{(t-1)}_n  > \eta \sigma^2 \right\}.
	\end{align}
	To satisfy the practical constraints introduced in \cite{nasir2019deep}, we limit the information exchange between AP $n$ and its neighboring APs as depicted in Fig. \ref{fig:infoexchangeconf}.
	Although, it is assumed in \cite{nasir2019deep} that receiver $n$ may do a more recent received power measurement from AP $i \in I^{(t+1)}_n$ just before the beginning of time slot $t+1$, i.e., $\bar{g}^{(t+1)}_{b^{(t)}_i\to n}p^{(t)}_i$, we prefer not to require it for our new model that involves mobility. Note that as device $n$'s association changes, i.e., $b^{(t+1)}_n \neq b^{(t)}_n$, we assume that the neighboring sets are still determined with respect to the previous positioning of AP $n$ and the feedback history from past neighbors is preserved at the new AP position to be used in agent $n$'s state. We let the association of device change only after staying within a new cell for $T_\textrm{register}$ consecutive slots.
	
	For the training process, as a major modification on \cite{nasir2019deep}, we introduce training episodes where we execute a training process for $T_\textrm{train}$ slots and let devices do random walk without any training for $T_\textrm{travel}$ slots before the next training episode. The $T_\textrm{travel}$ slot-long traveling period induces change in the channel conditions, and consequently allows policy to observe more variant states during its training which intuitively increases its robustness to the changes in channel conditions. To train a policy from scratch for a random wireless network initialization, we run $E$ training episodes which are indexed by $\mathcal{E}={1,\dots,E}$. The $e$-th training episode starts at slot $t_e=(e-1)\left(T_\textrm{train}+T_\textrm{travel}\right)$ and is composed of two parallel procedures called centralized training and distributed execution. After an interaction with the environment, each agent sends its newly acquired experience to the centralized trainer which executes the centralized training process. The trainer clears its experience-replay memory at the beginning of each training episode. Due to the backhaul delay as shown in Fig. \ref{fig:ddpgdiagram}, we assume that the most-recent experience that the trainer received from agent $n$ at slot $t$ is $e_n^{(t-1)}$. During slot $t$, after acquiring all recent experiences in the memory $\mathcal{D}$, the trainer runs one gradient step for the actor and critic networks. Since the purpose of the critic network is to guide the actor network during training, only the actor network needs to be broadcasted to the network agents and during the inference mode only the actor network is required. The trainer starts to broadcast $\bm{\theta}_\textrm{broadcast} \shortleftarrow \bm{\theta}_\textrm{agent}$ once every $T_u$ slots and we assume $\bm{\theta}_\textrm{broadcast}$ is received by the agents after $T_d$ slots again due to delay. In addition, compared to the deep Q-network in \cite{nasir2019deep} that reserves an output port for each discrete action, each actor network has just one output port.

	The local state of agent $n$ at time $t$, i.e., $s_n^{(t)}$ is a tuple of local environment features that are significantly affected by the agent's and its neighbor's actions. As described in \cite{nasir2019deep}, the state set design is a combination of three feature groups. The first feature group is called ``local information'' and occupies six neural network input ports. The first input port is agent $n$'s latest transmit power $p^{(t-1)}_n$ which is followed by its contribution to the network objective \eqref{eq:DynOptProblem}, i.e., $C^{(t-1)}_n$. Next, agent $n$ appends the last two measurements of its direct downlink channel and sum interference-plus-noise power at receiver $n$: $g^{(t)}_{n \to n}$, $g^{(t-1)}_{n\to n}$, $\left(\sum_{m \in \mathcal{N}, m \neq n}g^{(t-1)}_{m\to n}p^{(t-1)}_m + \sigma^2\right)$, and $\left(\sum_{m \in \mathcal{N}, m \neq i}g^{(t-2)}_{m\to n}p^{(t-2)}_m + \sigma^2\right)$.
	
	These are followed by the ``interfering neighbors'' feature group. Since we are concerned by the scalability, we limit the number of interfering neighbors the algorithm involves to $c$ by prioritizing elements of $I^{(t)}_{n}$ by their amount of interference at receiver $n$, i.e., $g^{(t-1)}_{i\to n}p^{(t-1)}_i$. We form $\bar{I}^{(t)}_{n}$ by taking first $c$ sorted elements of $I^{(t)}_{n}$. As $|I^{(t)}_{n}|<c$, we fill this shortage by using virtual neighbors with zero downlink and interfering channel gains. We also set its spectral efficiency to an arbitrary negative number. Hence, a virtual neighbor is just a placeholder that ineffectively fills neural network inputs. Next, for each $i\in\bar{I}^{(t)}_{n}$, we reserve three input ports: $g^{(t)}_{i\to n}p^{(t-1)}_i$, $C^{(t-1)}_i$. This makes a total of $3c$ input ports used for current interfering neighbors. In addition, agent $n$ also includes the history of interfering neighbors and appends $3c$ inputs using $\bar{I}^{(t-1)}_{n}$.
	
	Finally, we have the ``interfered neighbors'' feature group. If agent $n$ does not transmit during slot $t-1$, $O_n^{(t)} = \emptyset$ and there will be no useful interfered neighbor information to build $s_n^{(t)}$. Hence, we define time slot $t'_n$ as the last slot with $p_n^{(t'_n)}>0$ and we consider $O^{(t'_n+1)}_{n}$ in our state set design. We also assume that as agent $n$ becomes inactive, it will still carry on its information exchange between each $o \in O^{(t'_n+1)}_{n}$ without the knowledge of $g^{(t-1)}_{n\rightarrow o}$. Similar to the scheme described above, agent $i$ regulates $O^{(t'_n+1)}_{n}$ to set $|\bar{O}^{(t)}_{n}|=c$. For $o\in O^{(t'_n+1)}_{n}$, the prioritization criteria is now agent $i$'s share on the interference at receiver $o$, i.e., ${g^{(t-1)}_{n\to o}p^{(t-1)}_n}\left(\sum_{m \in \mathcal{N}, m \neq o}g^{(t-1)}_{m\to o}p^{(t-1)}_m + \sigma^2\right)^{-1}$. For each interfered neighbor $o \in O^{(t'_n+1)}_{n}$, $s_n^{(t)}$ accommodates four features which can be listed as: $g^{(t-1)}_{o\rightarrow o}$, $C^{(t-1)}_o$, and 
	${g^{(t'_i)}_{n\rightarrow o}p^{(t'_i)}_n}\left(\sum_{m \in \mathcal{N}, m \neq o}g^{(t-1)}_{m\rightarrow o}p^{(t-1)}_m + \sigma^2\right)^{-1}$.

	The reward of agent $n$, $r_n^{(t+1)}$, is computed by the centralized trainer and used in the training process. Similar to \cite{nasir2019deep}, $r_n^{(t)}$ is defined as agent's contribution on the objective \eqref{eq:DynOptProblem}:
	\begin{align}\label{eq:reward}
	\begin{split}
	r^{(t+1)}_n &= C^{(t)}_n - \sum_{o \in O^{(t+1)}_{n}} \pi^{(t)}_{n\rightarrow o}
	\end{split}
	\end{align}
	with $\pi^{(t)}_{n\rightarrow o} = \log\left(1+\gamma^{(t)}_{o}\left(\bm{b}^{(t)},\left[\dots,p^{(t)}_{n-1},0,p^{(t)}_{n+1},\dots\right]^\intercal\right)\right) - C^{(t)}_o$ being the externality that  link $n$ causes to interfered $o$.
	\section{Simulations}\label{sec:simulations}
\begin{figure}
		[t]
		\centering
		\includegraphics[clip, trim=0.40cm 0.3cm 0.40cm 0.40cm,width=0.8\columnwidth]{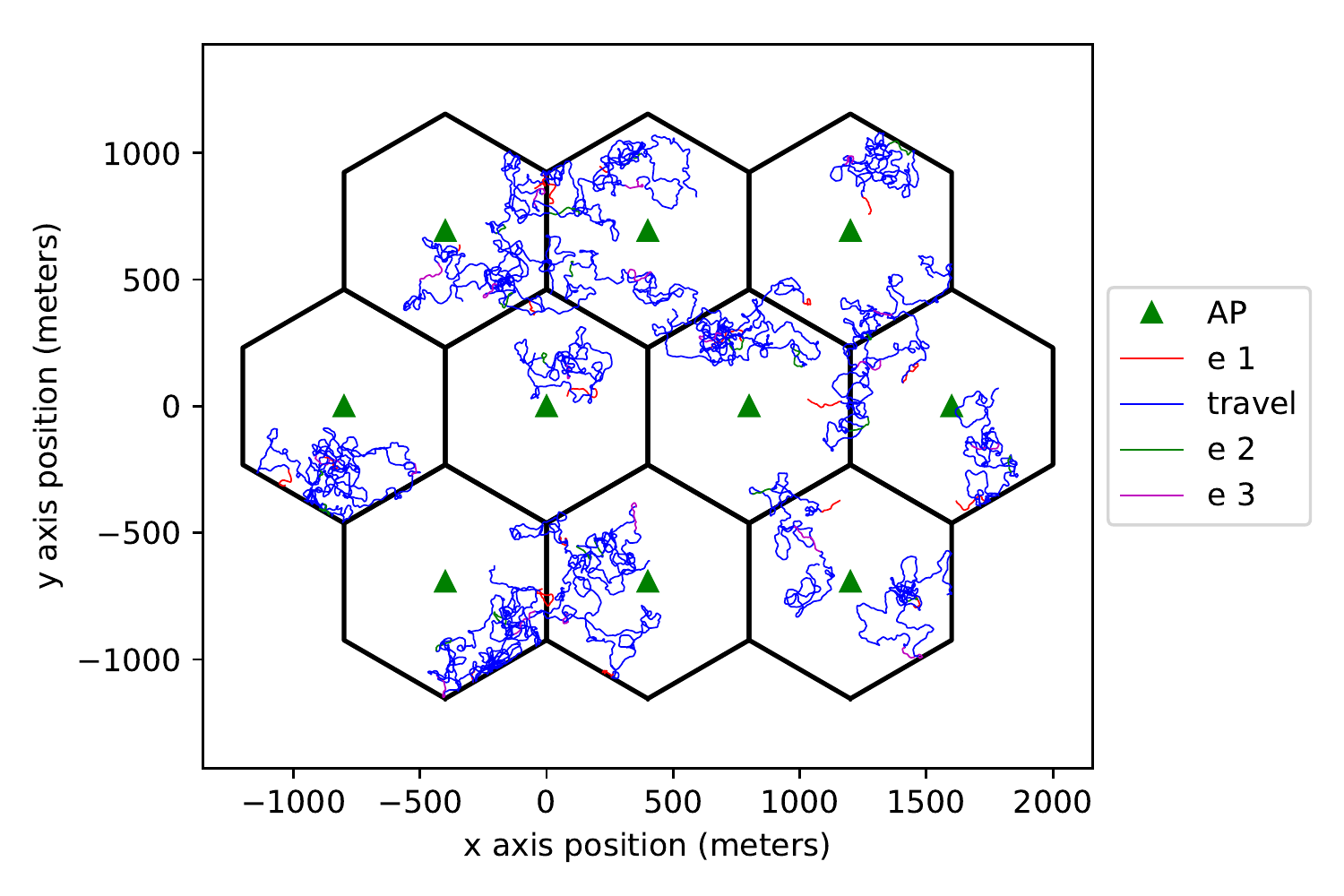}
		\caption{Example movement until the end of episode $e=3$.}
		\label{fig:movement}
	\end{figure}

	Following the LTE standard, the path-loss is simulated by $128.1 + 37.6\log_{10}(d)$ (in dB) with $f_c = 2$ GHz, where $d$ is transmitter-to-receiver distance in km. We set $\sigma_{\textrm{s}}=10$ dB, $d_{\textrm{cor}}=10$ meters, $T = 20$ ms, $P_{\textrm{max}}= 38$ dBm, and $\sigma^2=-114$ dBm. We simulate the mobility using Haas' model \cite{haas1997mobility} with maximum speed being $2.5$ m/s. Each mobile device randomly updates its speed and direction every second uniformly within $[-0.5,0.5]$ m/s and $[-0.175,0.175]$ radians, respectively. Fig. \ref{fig:movement} shows an example movement scenario until the end of third training episode with $T_\textrm{train} = 5,000$ and $T_\textrm{travel} = 50,000$ slots. The DDPG implementation and parameters are included in the source code. \footnote{GitHub repository: \url{https://github.com/sinannasir/Power-Control-asilomar}} Both WMMSE and FP start from a full power allocation, since it gives better performance than random initialization. WMMSE takes more iterations to converge than FP, resulting in higher sum-rate.
	\begin{figure}
		[t]
		\centering
		\includegraphics[clip, trim=0.0cm 0.25cm 0.25cm 0.30cm,width=1.0\columnwidth]{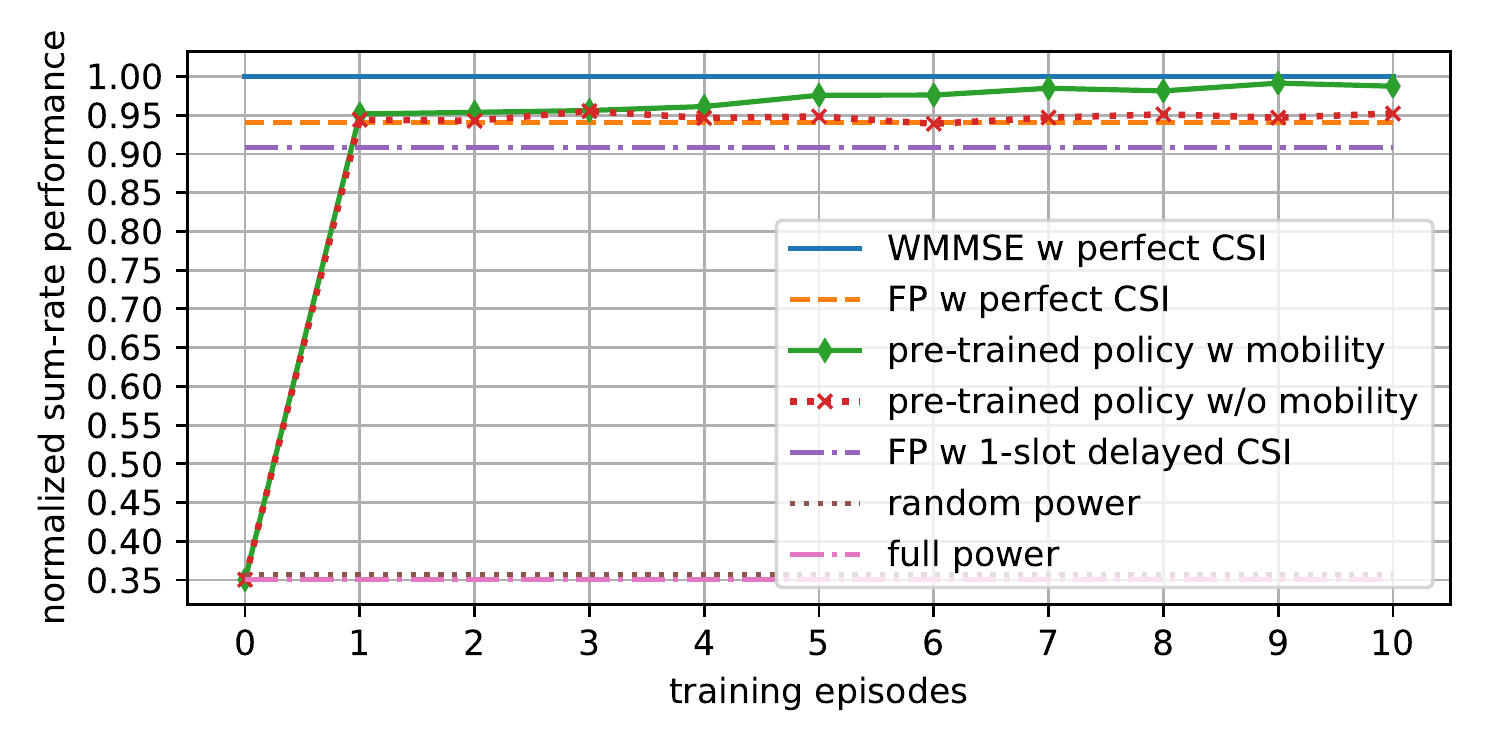}
		\caption{Test results for the 10 cells and 20 links scenario.}
		\label{fig:result}
	\end{figure}
	\begin{table}[t]
		\footnotesize
		\tabcolsep 0pt \caption{Average sum-rate performance in bps/Hz per link.}
		\begin{center}
			\def\temptablewidth{1\columnwidth}
			{\rule{\temptablewidth}{1pt}}
			\begin{tabular*}{\temptablewidth}{@{\extracolsep{\fill}}|c|c|c|c|c|c|c|}
				{(cells,links)} &policy trained for (10,20) & WMMSE & FP & FP w delay & random & full
				\\\hline \hline
				(10,20) &  2.59 & 2.61 & 2.45 & 2.37 & 0.93 & 0.91 \\
				(20,40) &  1.97 & 2.09 & 1.98 & 1.87 & 0.68 & 0.68 \\
				(20,60) &  1.58 & 1.68 & 1.59 & 1.50 & 0.37 & 0.35 \\
				(20,100)&  1.14 & 1.23 & 1.15 & 1.09 & 0.18 & 0.17 \\
			\end{tabular*}
			{\rule{\temptablewidth}{1pt}}
		\end{center}
		\label{table:scalability}
	\end{table}

	We first train two policies for $K=10$ cells and $N=20$ links network deployment for $E=10$ training episodes. The first policy is trained with mobile devices, whereas the latter is trained without mobility, i.e., with steady channel. We set $f_d$ to 10 Hz for all time slots \cite{nasir2019deep}. We save the policy parameters during training for testing on several random deployments with $(K,N) = (10,20)$ and mobility. As shown in Fig. \ref{fig:result}, without mobility, there is no significant sum-rate gain after the first training episode and policy converges to FP's sum-rate performance. As a remark, FP is centralized and it has full CSI, whereas actor network is distributively executed with limited information exchange. As we include device mobility and a certain travel time between training episodes, the policy is able to experience various device positions and interference conditions during training, so its sum-rate performance consistently increases. Additionally, in Table \ref{table:scalability}, we show that an actor network trained for $(K,N) = (10,20)$ can keep up with the sum-rate performance of optimization algorithms as network gets larger. Hence, running centralized training from scratch is not necessary as device positions change or new devices register, since a pre-trained policy for a smaller and different deployment performs quite well. For the 20 link scenario, on average, WMMSE and FP converge in 42 and 24 iterations, respectively. For 100 links, WMMSE requires 74 iterations. Conversely, learning agent takes just one policy evaluation. 
	
	\section{Conclusion}\label{sec:conclusion}
	In this paper, we presented a distributively executed deep actor-critic framework for power control. During training, only actor network is broadcasted to learning agents. Simulations show that a pre-trained policy gives comparable performance with WMMSE and FP, and a policy trained for a smaller deployment is applicable to a larger network without additional training thanks to the distributed execution scheme. Further, we have shown that the proposed actor-critic framework enables real-time power control under certain practical constraints and it is compatible with the case of mobile devices. DDPG in fact uses the mobility to increase its sum-rate performance by experiencing more variant channel conditions.
	\bibliographystyle{IEEEtran}
	\bibliography{ref}{}
\end{document}